\documentstyle[aps,pre]{revtex}

\begin{document}

\title{Influence of boundary conditions on the dynamics of pattern
forming systems: the case of oscillatory media.}

\author{S. Bouzat \thanks{Electronic address:
bouzat@cab.cnea.gov.ar; Fellow of CONICET, Argentina} and H. S.
Wio
\thanks{Electronic address: wio@cab.cnea.gov.ar; Member of
CONICET, Argentina}}

\address { Grupo de F\'{\i}sica Estad\'{\i}stica
\thanks{http://cab.cnea.gov.ar/cab/invBasica/fisEstad/estadis.htm}
\\ Centro At\'omico Bariloche (CNEA) and Instituto Balseiro (CNEA
and UNC),\\ 8400-San Carlos de Bariloche, Argentina}

\maketitle

\begin{abstract}
We study the pattern dynamics in a reaction diffusion model of the
activator--inhibitor type in the oscillatory regime. We consider
finite systems with partially absorptive boundary conditions
analizing examples in different geometries in one and two
dimensions. We observe that the boundary conditions have
important effects in the pattern forming properties of the
systems. In all the studied cases, the arising complex behaviour
is found te be dependent on the absorption parameter. By changing
this parameter we can control the asymptotic behaviour to be
stationary, periodic, quasiperiodic or chaotic.
\end{abstract}

%\pacs{05.45.-a,47.54.+r,47.52.+j}
%\keywords{Reaction--diffusion, boundary conditions, spatiotemporal chaos.}

\section{\bf Introduction}

There are many properties in pattern forming reaction--diffusion
systems that do not depend on the geometry of the system (for
instance Turing wavelengths, frequencies of oscillation, shapes
and velocities of waves or fronts.) \cite{general}. Due to this
fact, most studies on pattern formation have not paid too much
attention to features such as the shapes and sizes of the reactors or
the boundary conditions, working mainly with infinite systems or
periodic boundary conditions. However, it is well known that
boundaries and boundary conditions can play a relevant role in
determining the behaviour of this kind of systems. For instance,
in reference \cite{ramped} the importance of geometrical
characteristics is quiet remarkable. Also, recent theoretical papers
have shown that the domain shape may have an important effect in
the propagation of waves in excitable media \cite{piro1}, and
that the consideration of non trivial boundary conditions
\cite{piro2} in some reaction diffusion--systems leads to
qualitatively different dynamic behaviour from the one obtained in
infinite media.

Here we analyze problems where the boundary conditions
(bc's), as well as the size of the system, are relevant parameters
determining the kind of complex dynamics arising. We study
an activator--inhibitor model in the oscillatory regime
considering different geometries with partially absorptive bc's.
For a density field $\rho(\vec{r},t)$ of a given reactant this
condition is written as
\begin{equation}
- D (\nabla \rho)|_\sigma \cdot \vec{n} = k  \rho |_\sigma,
\end{equation}
where $D$ is the diffusion constant, the subscript $\sigma$
indicates evaluation at the boundary, $\vec{n}$ is the unit
vector normal to the boundary in the outer direction, and $k(\ge 0)$
is the absorption or albedo parameter. This relation expresses
the fact that there is an outgoing flux of reactants through the
boundary which is proportional to the density. The limit $k=0$
corresponds to no flux or Neumann bc's, which means a perfectly
reflecting (non absorbing) boundary, whereas clearly, an increasing
value of $k$ corresponds to an increasing absorption at the
boundary, and a reduced reflectivity.

We consider the reaction--diffusion model given by the equations
\begin{eqnarray}
\label{sys}
\dot{u}&=&\nabla^2 u+f(u)-v          \nonumber \\
\dot{v}&=&D_v \nabla^2 v + u -\gamma v - c,
\end{eqnarray}
where the non linear function $f(u)=-(u-u_0)^3+u$ characterizes
the autocatalytical properties of the activator $u$. The model
shares the general features of more realistic ones describing
chemical systems such as the Belousov Zabotinskii or CIMA reactions
\cite{tyson,LE}. For $u_0=c=0$, it was used in the analysis of
the propagation of fronts in bistable systems \cite{meron} and
also in the study of pattern formation in inhomogeneous media
\cite{PLAinh,PREinh}. The constant $c$ is here set equal to $[u_0
(1-\gamma)]$. With this election $u_0$ represents a translation
of the related spatially independent system (i.e. Eqs.
(\ref{sys}) without diffusion terms) in the ($u,v$)-plane along
the line $(u=v)$. For $\gamma>1$ the equations describe a
bistable medium, while for $\gamma<1$ they describe an
oscillatory one. Here, we consider $\gamma=.9$. In order to work
with positive values of the densities $u$ and $v$, we set
$u_0=.4$. This last fact is desirable taking into account the
kind of bc's we will work with, however, it is not strictly
necessary, as we have found the same kind of results (in all the
problems here discussed) using $u_0=0$.

For the chosen values of parameters, the homogeneous system
associated to Eqs. (\ref{sys}) has a limit cycle solution of
period $\tau_0\simeq 14.66$. As initial conditions for our
extended system we consider homogeneous states belonging to this
limit cycle. This (natural) choice is taken in order to focus our
attention on the effects caused by the boundary conditions. The
consideration of inhomogeneous initial states would produce
additional pattern forming phenomena that would interact with
those coming from the boundary conditions. An important
characteristic of the extended system is that, for $D_v>1,73$, a
Turing instability \cite{general,Nicolis} of wavelength
$\lambda_T\simeq13$ coexists (and competes) with the Hopf
instability \cite{general,Nicolis}.

The aim of this paper is to show that a wide spectrum of
possibilities concerning pattern dynamics arises when different
boundary conditions are considered in an oscillatory medium. We
would like to attract the interest of experimentalists to study
these situations. The experimental observation of the phenomena
here predicted might be realized by adequately designing chemical
systems sharing the properties of our models. Also, the partially
absorptive conditions may appear in real systems as defects at the
boundaries, such as, small separations between the gel and the
wall of the reactor containing the solution with uncontrolled
concentration of reactants. Another experimental possibility is to
work with electrical systems \cite{Pur}.

All numerical calculations were done as follows. Firstly, the
different systems of partial differential equations have been
approximated by systems of coupled ordinary differential
equations, obtained by finite difference schemes. Secondly the
resulting equations have been solved by a Runge--Kutta 2 method.
Different space and time discretization schemes were employed in
order to check the results.

The organization of the paper is as follows: in Section II we
study a unidimensional system with Neumann boundary conditions at
one border and partially absorptive at the other. In Section III we
study two different bidimensional systems corresponding to a
circular reactor with partially absorptive boundary conditions
and a long rectangular system with different boundary conditions
on its walls. The last Section is devoted to discussing the
results and drawing some conclusions.

\section{Unidimensional system.}

We now consider Eqs.(\ref{sys}) in the one dimensional domain
$0\le x \le L$, with Neumann bc's at $x=0$ and partially
absorptive bc's at $x=L$:
\begin{eqnarray}
\partial_x u(0)&=&0 \,\,\,\,\,\,\,\,\,\,\,\,\,\,\,\,\,\,\,\,\,\,\,\,\,\,\,\,
 D_v \partial_x v(0)=0    \nonumber \\
\partial_x u(L)&=&-k_u u(L) \,\,\,\,\,\,\, D_v \partial_x v(L)=-k_v v(L).
\end{eqnarray}
As initial conditions we consider homogeneous states belonging to
the limit cycle of the associated non extended system (Eqs.
(\ref{sys}) without diffusion). We analyze the dynamics for
different values of $L$,$D_v$, $k_u$ and $k_v$.

At short times, near $x=0$, where there are Neumann bc's, the
system evolves very close to the natural homogeneous limit cycle
of the medium; whereas, near $x=L$, the motion is perturbed by
the albedo bc's which cause the local oscillations to be slower
and of smaller amplitude. This effect increases with the
absorption parameter. Hence, the dynamics of the system begins
to be inhomogeneous and the perturbation is transmitted from
$x=L$ to the rest of the system, which asymptotically reaches
different dynamical regimes depending on the parameters. The
resulting regime is independent of the initial condition (when
taken on the limit cycle as mentioned above) except maybe in some
specific parameter regions, as we will discuss later.

For very high absorption, the oscillations near $x=L$ are
inhibited. As a consequence of this, for small enough systems
(typically $L<2 \lambda_T$, but depending on $D_v$ and $k$'s), the
oscillations are found to be stopped in the whole system and
stationary patterns are generated. This is because the stationary
tendency of the zone near $x=L$ affects the whole system. The
effect is increased when $D_v$ grows, since the coupling of the
rest of the system to the indicated zone becomes stronger. (For a
higher value of $D_v$, a stationary pattern may appear for lower
values of the absorption parameter or in a larger system.)

In general, when this stationary regime does not occur, the
oscillations are not inhibited. The system asymptotically reaches
periodic, quasiperiodic, or chaotic inhomogeneous regimes.
However, for large enough $D_v$ (typically $D_v>2.3$), the system
evolves toward a Turing pattern, that is also stationary, but of a
completely different origin \cite{Nicolis} than the stationary
patterns described above. Turing patterns are generated by means
of freezing fronts \cite{Pur} which travel from $x=L$ (the
slower oscillating zone) to the rest of the system. As they
advance, they inhibit the temporal oscillations and leave behind
the characteristic spatially--periodic structures. The process of
formation of a Turing pattern is shown in Fig.1.

In Fig. 2 we show the phase diagram indicating the asymptotic
behaviour of a system of length $L=70\simeq 5 \lambda_T$ with
$k_u=k_v\equiv k$ as function of $D_v$ and $k$. The different
regions correspond to Turing patterns (TP), inhomogeneous periodic
oscillations (PO), inhomogeneous quasiperiodic oscillations (QP)
and spatiotemporal chaos (CH).

In the periodic region (PO), the asymptotic behaviours correspond
to regimes of periodic traveling waves from $x=L$ (albedo bc) to
$x=0$ (Neumann bc). The global period of the motion results
always a little bit larger than $\tau_0$. This is because of the
early slowdown of the oscillations occurring near $x=L$, that is
later transmitted to the whole system. In Fig. 3 we show a
spatiotemporal plot of the $u$--field corresponding to a time
window of a typical asymptotic regime in the PO region.

To distinguish between PO, QP, and CH regimes we consider the
succession $\{t_n\}$ of times at which $u(x=0,t)$ reaches a local
maximum as function of $t$, i.e. the times when
\begin{equation}
\frac{\partial }{\partial t} u(0,t)|_{t_n} = 0  \,\,\,\,\,\,\,
{\rm and}  \,\,\,\,\,\,\, \frac{\partial^2 }{\partial t^2} u(0,t)|_{t_n} < 0 \nonumber
\end{equation}
occur simultaneously. Then we define a new succession
$\{p_n=t_n-t_{n-1}\}$.

The study of the behaviour of $\{p_n\}$ provides a useful and
simple way to determine the character of the dynamics (periodic,
quasiperiodic or chaotic), as was already observed when dealing
with inhomogeneous systems \cite{PREinh}. In the case of periodic
motion the value of $p_n$ converges to a constant $p_\infty$ which
coincides with the period of the global motion. In the case of
quasiperiodic motion the $p_n's$ asymptotically show periodic
or quasiperiodic behaviour. In the chaotic region the $p_n's$
exhibit highly disordered behaviour. Instead of considering the
succession $\{p_n\}$, we work with the values of $p_n$ plotted as
a function of $t_n$. Hence we define a function of time
$p(t)\equiv p_n$ (strictly defined only for the discrete values
$t=t_n$). In Fig. 4 we show the typical plots of the $p(t)$ for
the PO, QP and CH regimes. In Fig. 4.a. the early time behaviour
can be observed, in which the system at $x=0$ is performing
the natural limit cycle oscillation of the medium, until the
perturbation coming from $x=L$ arrives slowing sown the motion. It
is worth here mentioning that we have not observed marginal
profiles of $p(t)$ where the determination of the kind of
dynamics was not clear. However, in the transition zones
between phases, the transients are long and it is necessary to
observe the evolution for a very long time. Another remarkable
fact is that disordered behaviour such as that appearing in
Fig. 4.c. -here called chaotic- are chaotic in the sense of
showing high sensitivity to initial conditions. We have verified
this fact in a large number of cases by studying the evolution of
the distance between solutions with different (but close) initial
conditions, in the same way as it was done in \cite{PLAinh}.

In the asymptotic regimes of the QP and CH regions, it is not only waves
traveling from $x=L$ to $x=0$ that appear. There are intervals during
which waves travel in the opposite sense, and also, from time to time, a
center emitting waves in both directions may appear in any part of
the system (specially in the CH regime). These effects are
observed as defects and dislocations in the spatiotemporal plots
of the $u$--fields. In the QP region, the defects appear
periodically or quasiperiodically in time, while in the CH region
they appear in a random way. In Fig. 5 we show a spatiotemporal
plot corresponding to a chaotic regime, together with the plot of
$p(t)$ for the same time window. There, it can be observed that
the arising of defects is related to the occurrence of maxima in
$p(t)$, which indicates delays in the oscillations at $x=0$.

In a small region of the transition zone between the PO and TP
regions, which appears shadowed in the diagram of Fig. 2, the
behaviour of the system is complicated. In general, a long chaotic
transient appears that, later, eventually converges to Turing
patterns, periodic or quasiperiodic behaviour. In this zone, the
asymptotic behaviour is also possibly dependent on the initial
condition -even when homogeneous and belonging to the
indicated limit cycle. We have observed also Turing patterns that
are generated by freezing fronts coming from $x=0$ (the Neumann
bc) instead of from $x=L$ (the albedo bc). One possible
explanation for such peculiarities is that, in this zone of the
diagram, the Turing instability is well established, but the
albedo bc's introduce only a very small perturbation (because of
the small values of $k$s) that it is not always strong enough
to excite the unstable spatial mode.

In the phase diagram of Fig. 2 we have indicated with dotted and
dashed lines the zones where the transition between periodic and
nonperiodic regimes respectively occur for a system with
($k_u=0,k_v\equiv k$) and ($k_u\equiv k,k_v=0$). It can be seen
that for ($k_u=0,k_v\equiv k$) the transition occurs for similar
--or even lower-- values of $k$ than in the case $k_u=k_v=k$ but,
for ($k_u\equiv k,k_v=0$), much larger values of $k$ are needed
in order to observe nonperiodic regimes. This indicates that the
system is more sensitive to changes (or imperfections) in the
boundary conditions of the $v$--field than of the $u$--field.

In the next section we analyze the dynamics of Eqs.
(\ref{sys}) with partially absorptive bc's in two different
bidimensional geometries.

\section{Bidimensional systems.}

\subsection*{Circular system.}
Firstly we consider the activator inhibitor model of
Eqs.(\ref{sys}) in a circular system of radius $R$ with albedo bc
for both fields
\begin{equation}
\partial_r u(R)=-k_u u(R), \,\,\,\,\,\,\,\,\,\,\,\,\,\, D_v \partial_r v(R)=-k_v v(R).
\end{equation}
where $r^2=x^2+y^2$ is the radial coordinate. We impose
homogeneous initial conditions as in the previous section and
study numerically the radial equations corresponding to
(\ref{sys}):
\begin{eqnarray}
\label{sys2}
\dot{u}&=& \frac{1}{r} \frac{\partial}{\partial r}(r\frac{\partial u}{\partial r}) +f(u)-v            \nonumber \\
\dot{v}&=&D_v \frac{1}{r} \frac{\partial}{\partial r}(r\frac{\partial v}{\partial r}) + u -\gamma v -c,
\end{eqnarray}
with $f(u)$, $c$, $\gamma$ and $u_0$ as before. Analyzing the
spatiotemporal plots $(r,t)$ of the solutions and the $p_n$s (now
defined at $r=0$), we have observed the same kinds of behaviour
(periodic, quasiperiodic chaotic and Turing) as in the
unidimensional system of the last section. In Fig. 6 we show the
phase diagram for two different values of the radius of the
system $R$.

In the first case, for $R=70$, there is no CH phase because the
system is not large enough to allow the required disorder on the
dynamics of the oscillating phase in the different parts of the
system to be produced. However, some chaotic isolated points have
been observed in the shadowed zone in the transition to the TP
region, where the properties of the dynamics are similar to those
explained for the one--dimensional problem. For $R=80$ we have the
same phases as in the unidimensional problem. Here, the
asymptotic regimes correspond to target patterns, that are
stationary in the TP region, and of waves traveling from $r=R$ to
$r=0$ in the PO region, and traveling in both directions in the
QP and CH regions.

In the CH region, the radial symmetry is expected to be broken by
small angular perturbations in real systems.

In both diagrams we show the transition lines from periodic to
nonperiodic regimes for the cases ($k_u\equiv k,k_v=0$) and
($k_u=0,k_v\equiv k$). The observed shift of these lines relative
respect to that for $k_u=k_v=k$ are similar to those in the
one dimensional problem. The conclusion that the system is more
sensitive to boundary conditions on the $v$--field remains valid
in bidimensional systems.

\subsection*{Band shaped system.}
Now we consider Eqs.(\ref{sys}) in the same parameter region as
before, in a long rectangular domain ($0 \le x \le L_x, -L_y \le
y \le L_y$) with $L_y<<L_x$ with different boundary conditions on
its sides. We fix albedo bc at $x=0$ with $k_u=k_v=10$ (which
means a strong absorption or very small reflectivity) and Neumann
bc at $x=L=80$ ($k_u=k_v=0$). If we now fix Neumann bc also at
the $y=\pm L_y$ sides, we will have (for homogeneous initial
conditions) an effective one--dimensional system equivalent to the
one analyzed in section II. Instead, here we consider another
albedo bc at $y=\pm L_y$ with a different value of $k\equiv
k_u=k_v$. In this case the patterns will not have translational
symmetry in the $y$--direction. We fix $D_v=2$ and study
numerically the solutions for different values of $k$ and $L_y$.
The situation to analyze is sketched in Fig. 7.

The one dimensional problem with $k=10$ at one end, and $k=0$
at the other is chaotic for $L=80, D_v=2$. Now, in the
bidimensional system considered, it is expected that when
increasing the albedo  parameter on the $y=\pm L_y$ sides from
zero, the dynamics will begin to differ from the one of the
one--dimensional system. In particular, if $L_y$ is small and $k$
large, we expect that the system will become stationary because of
the stationary tendency introduced by this new boundary condition
that will constrain the oscillations from the sides. This is in
fact what simulations have shown. There are also intermediate
values of $k$ and $L_y$ where the stationary tendency of the
sides is not enough to completely inhibit the oscillations, but
it is enough to change their characteristics from chaotic to
quasiperiodic or periodic.

In Fig. 8 we show a diagram of the $(k,L_y)$--plane indicating
the stationary and non--stationary regions as well as some points
where periodic, quasiperiodic and chaotic dynamics have been
observed.

It is posible to use a simple theoretical scheme in order to
explain qualitatively the transition of the system from non
stationary regimes to stationary ones when $k$ is increased
or $L_y$ decreased. Such a scheme is based on a concept
originated within the framework of nuclear reactor physics that
simplifies noticeably the analysis of the so called criticality
problems or neutron density distributions, in many situations of
interest. This is the concept of {\it geometrical buckling}.
It helped to reduce the dimensionality of the problem,
by assuming that directions transverse to the more relevant one can
be essentially described by the fundamental mode. Such an
assumption allows to include the effect of those extra dimensions
through (lateral) "sink" contributions (that is taking into
account the lateral leak) in a (typically one--dimensional)
diffusion equation for the neutron density on the relevant
direction \cite{reactor}. For instance, using this approximation
it is possible to estimate the critical mass or radius of a
nuclear system. Clearly, such an approximation gives reliable
results due to the linearity  of the neutron diffusion equation
(at least over time scales shorter than those at which the transport
coefficients --cross section, diffusion constants-- change).

Our system of equations is clearly nonlinear, hence such an
approach could in principle not be applied. However, we
qualitatively introduce such an idea forcing our equations to reduce
their dimensionality and including associated sink or leak terms.

For the system in Eq.(\ref{sys}) in the rectangular geometry with
$L_y<<L_x$, and albedo conditions at $y=\pm L_y$ given by
\begin{equation}
\label{albfr}
\partial_y u(x\pm,L_y)=\mp k u(x,L_y), \,\,\,\,\,\,\,\,\,\,\,\,\,\,
D_v \partial_y v(x,\pm L_y)=\mp k v(x,L_y),
\end{equation}
we will assume that the solution can be written as
\begin{eqnarray}
u(x,y,t)&=&\tilde{u}(x,t) \cos(q_u y) \nonumber \\
v(x,y,t)&=&\tilde{v}(x,t) \cos(q_v y).
\end{eqnarray}
Hence, we are approximating the profiles of the distributions of
$u$ and $v$ in the $y$ coordinate as being independent of $x$ and
$t$ except for a global factor. The cosine functions give a more or
less desirable form for the densities, having a maxima at $y=0$.
Clearly, the ($k=0$)--case corresponds to $q_u=q_v=0$ with
solutions independent of $y$. We may expect that our approximation
will give qualitatively better results for small $k$. From Eq.
(\ref{albfr}) we find the following relation that determines
$q_u$ and $q_v$ as function of $k,L_y$ and $D_v$:
\begin{equation}
\label{tan}
k=q_u \tan(q_u L_y)  \,\,\,\,\,\,\,\,\,\,\ k=q_v D_v \tan(q_v L_y),
\end{equation}
and from Eq.(\ref{sys}) we obtain
\begin{eqnarray}
\dot{\tilde{u}}&=& \frac{\partial \tilde{u}}{\partial x^2}- q_u^2 \tilde{u} +
f(\tilde{u})-\tilde{v} \nonumber \\
\dot{\tilde{v}}&=& \frac{\partial \tilde{v}}{\partial x^2}- D_v q_v^2 \tilde{v} + \tilde{u} -
\gamma \tilde{v} - c.
\end{eqnarray}
Now, since $L_y<<L_x$, we might think that, whether the dynamics is
stationary or not will depend only on $k$ and $L_y$ and not on
what happens on the $x$ coordinate\cite{coment}, and we may
consider the system as independent of $x$. We then obtain the
following equations for a space independent activator--inhibitor
system:
\begin{eqnarray}
\dot{\tilde{u}}&=& f(\tilde{u})- q_u^2 \tilde{u}-\tilde{v} \nonumber \\
\dot{\tilde{v}}&=& \tilde{u}- D_v q_v^2 \tilde{v} -\gamma \tilde{v} - c.
\end{eqnarray}
This system, for the chosen values of $f(u)$, $\gamma$, and $c$,
is oscillatory for null or small values of $q_u$ and $q_v$, but
there exist critical values of these quantities (which depend on
$L_y$, $D_v$ and $k$ through Eqs.(\ref{tan})) beyond which the
system begins to be monostable, i.e. asymptotically stationary. We
associate this stationarity to that of our extended system. In
Fig. 8 we have plotted with a dotted line the limit from
stationary to non stationary regimes predicted in this way.
Although the results do not agree quantitatively with those
coming from simulations in the extended bidimensional system
(solid line), the occurrence of the phenomenon is clearly
well predicted qualitativelly.

\section{Final Remarks}

We have analyzed the influence of partially absorptive boundary
conditions in a typical reaction diffusion--model of an
oscillatory active medium with different geometries. We have
observed that the homogeneous periodic behaviour expected for the
case of Neumann boundary conditions, when homogeneous initial
condition are considered, is completely altered when the
absorption parameter is increased from zero. For small
absorption, the behaviour continues being periodic although the
homogeneity is lost and traveling waves are observed. For higher
absorption (with precise values of $k$'s depending on the
diffusion constants), the periodicity is broken and quasiperiodic
inhomogeneous oscillation and spatiotemporal chaos arise.

The kind of phenomena described here is expected to appear in real
systems sharing the properties of the model. The predictions made
for the three systems analyzed (one--dimensional,
circular and band--shaped) illustrates what can be
expected in more complicated situations. For example, let us
consider the same dynamics given by Eqs. (\ref{sys}) with
homogeneous initial conditions, but now in a square reactor with
Neumann bc's everywhere except in a sector where albedo bc's are imposed. In
Fig. 9 we show the contour plots of the $u$--fields for this
system at several times taken in the asymptotic regime,
assuming standard values of the parameters $D_v$ and $k$. The
spatiotemporal complexity arising is apparent.

When analyzing the transition lines from periodic to nonperiodic
regimes for the cases ($k_u\equiv k,k_v=0$) and ($k_u=0,k_v\equiv
k$) we have observed shifts of these lines relative to the
case $k_u=k_v=k$ that are similar in both the one dimensional and
the circular cases. These results imply that the system is more
sensitive to boundary conditions on the $v$--field than on the
$u$--field.

As mentioned in the introduction, in a real experiment on
oscillatory media, the waves emerging from the boundaries with
imperfect reflectivity will interact with other patterns that may
be spontaneously generated from fluctuations in the distribution
of reactants in any part of the system. In this paper we have
analyzed simpler situations that allowed us to isolate clearly
the effects of the boundaries. The results obtained constitute a
starting point for the study of more realistic situations.

\vspace{1cm}

{\bf ACKNOWLEDGMENTS:} The authors thank V. Grunfeld for a
revision of the manuscript. Partial support from CONICET (grant
PIP Nro.4593/96), from Argentine, is also acknowledged.

%1
\begin{figure}[tbp]
\caption{Propagation of a freezing front leading to the formation
of a Turing pattern. Spatiotemporal plot of the $u$--field for
$L=70$, $D_v=2.4$ and $k_u=k_v=1$.}
\end{figure}

%2
\begin{figure}[tbp]
\caption{Phase diagram (($D_v,k$)-plane) indicating the
asymptotic behaviour of the one dimensional system for $L=70$ and
$k_u=k_v\equiv k$. The regions corresponding to inhomogeneous
periodic oscillations, Turing patterns and quasiperiodic and chaotic
behaviour, are indicated with PO, TP, QP, CH respectively.
The transitions from periodic to nonperiodic regimes in the cases
$(k_u\equiv k, k_v=0)$ and $(k_u=0, k_v\equiv k)$ are indicated
respectively with dashed and dotted lines.}
\end{figure}

%3
\begin{figure}[tbp]
\caption{Spatiotemporal plots of the $u$--fields in the
asymptotic regime of the PO region. Calculations for $D_v=1.8,
k_u=k_v=.02$.}
\end{figure}

%4
\begin{figure}[tbp]
\caption{Typical plots of $p(t)$ in for: (a) PO region ($D_v=1.8,
ku=kv=.02$); (b) QP region ($D_v=1., k_u=k_v=1.$); (c) CH region
($D_v=1.8, k_u=k_v=1.$).}
\end{figure}

%5
\begin{figure}[tbp]
\caption{Dynamics in the CH region. Top: spatiotemporal plot of
the $u$--field for $D_v=1.8, k_u=k_v=1$ in a temporal window of the
asymptotic regime. Bottom: plot of $p(t)$ for the same system in
the same time window.}
\end{figure}

%6
\begin{figure}[tbp]
\caption{Phase diagrams indicating the asymptotic regimes of
circular systems for: (a) $L=70$; (b) $L=80$. In both cases, the
regions are indicated with the same nomenclature as in figure 2.
With dashed and dotted lines we indicate the transitions from
periodic to nonperiodic regimes in the cases $(k_u\equiv k,
k_v=0)$ and $(k_u=0, k_v\equiv k)$ respectively.}
\end{figure}

%7
\begin{figure}[tbp]
\caption{Sketch of the band shaped system indicating the
different boundary condition considered.}
\end{figure}

%8
\begin{figure}[tbp]
\caption{Diagram indicating the different behaviours observed in
the band shaped system as function of $k$ and $L_y$. The solid
line indicates the transition between stationary and nonstationary
regimes calculated from simulations. The dotted line corresponds
to the estimation of this frontier with the discussed theoretical
arguments. Crosses, squares, circles and triangles correspond to
the observations in simulations of stationary, periodic,
quasiperiodic and chaotic behaviour respectively.}
\end{figure}

%9
\begin{figure}[tbp]
\caption{Contour plots of the $u$-fields for a square system with
Neumann bc's except in the right bottom corner (indicated with
arrows in the left figure) where we have fixed albedo bc's with
$k_u=k_v=1$. From left to right the plots correspond to times
$t=2000,2004,2008,2012$ and $2016$. The other parameters are
$D_v=2$, and $L_x=L_y=60$.}
\end{figure}


\begin{thebibliography}{99}

\bibitem{general}  M.C.~Cross and P.C.Hohenberg, Rev. Mod. Phys.
{\bf 65}, 851 (1993); D. Walgraef, {\it Spatio-temporal pattern
formation}, (Springer-Verlag, New York , 1997).

\bibitem{ramped} E. Dulos, P. Davies, R. Rudovics and P. De
Kepper, Physica D {\bf 98} 53 (1996).

\bibitem{piro1} I. Sendi\~na--Nadal, V. P\'erez--Mu\~nuzuri, V. M.
Egu\'{\i}luz, E. Hern\'andez--Garc\'{\i}a and O. Piro, {\it
Quasiperiodic patterns in boundary-modulated excitable waves},
Phys. Rev. Lett. submited.

\bibitem{piro2} V. M. Egu\'{\i}luz, E. Hern\'andez--Garc\'{\i}a
and O. Piro, Physica A {\bf 283} 48 (2000).

\bibitem{tyson} J. J. Tyson and P. C. Fife, J. Chem. Phys. {\bf 73}
N.5 2224 (1980).

\bibitem{LE} Lengyel I. and Epstein I. R., Proc. Natl. Acad. Sci.
USA {\bf 89} 3977 (1992).

\bibitem{meron} A. Hagberg and E. Meron, Nonlinearity {\bf 7}
 805 (1994).

\bibitem{PLAinh} S. Bouzat and H.S. Wio, Phys. Lett. A {\bf 268} 323
(2000).

\bibitem{PREinh} S. Bouzat and H.S. Wio, {\it Pattern dynamics in
bidimensional oscillatory media with bistable inhomogeneities},
Phys. Rev. E (in press).

\bibitem{Nicolis} G. Nicolis, {\it Introduction to Nonlinear
Science}, (Cambridge U.P., Cambridge, 1995).

\bibitem{Pur} G. Heidemann, M. Bode and H.G. Purwins, Phys. Lett. A
{\bf 177} 225 (1993).

\bibitem{reactor} P.F. Zweifel, {\it Reactor Physics}, McGraw-Hill,
NY (1973), chp.3; K.H Beckurts and K. Wirtz, {\it Neutro Physics},
(Springer-Verlag, Berlin 1964), chp.9.

\bibitem{coment} The analysis of the dynamics of the $x$ coordinate
would determine, in case the motion is not stationary, whether it
is periodic, quasiperiodic, or chaotic; but this is an aspect in
which we are not interested here.

\end{thebibliography}
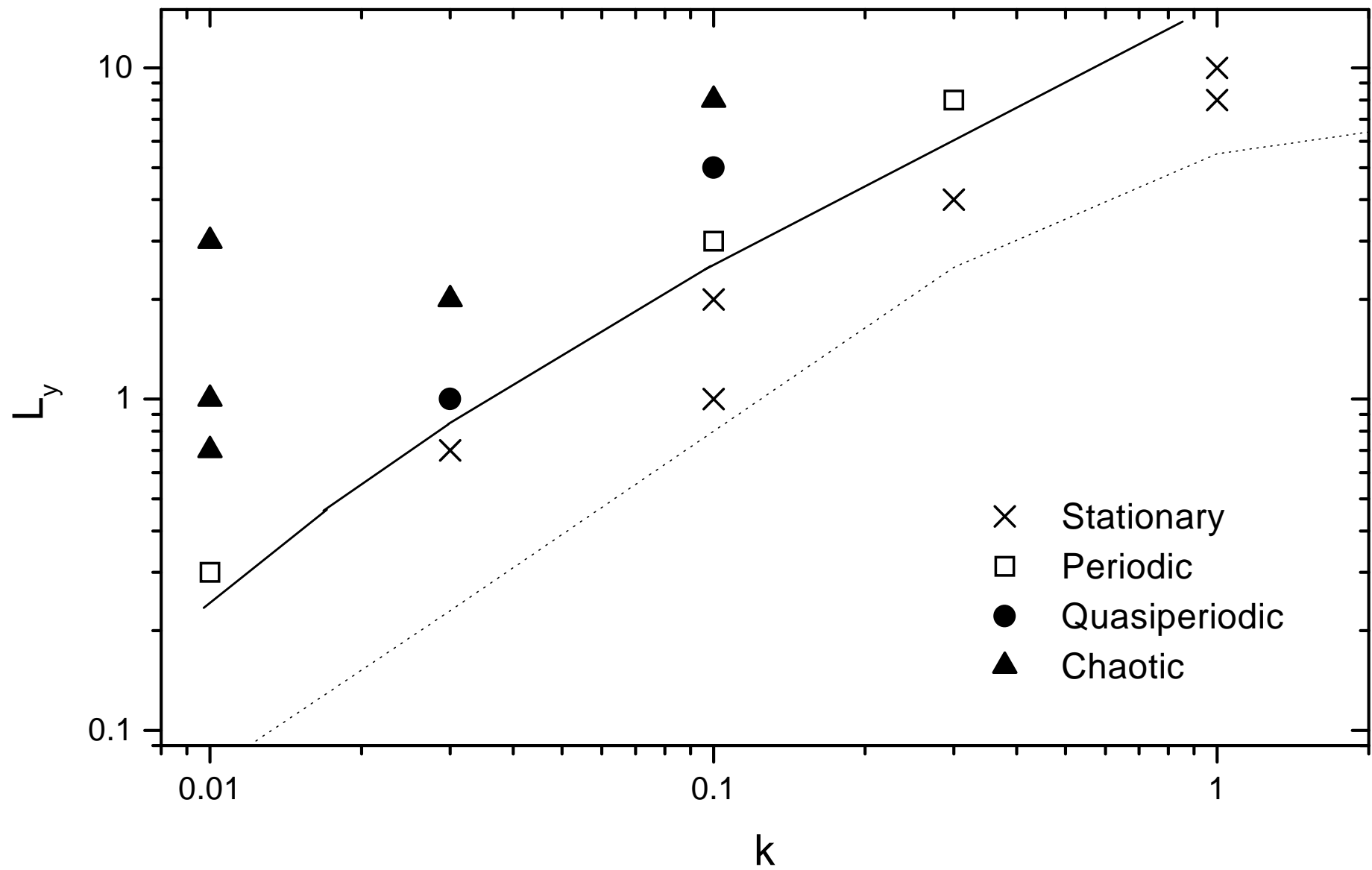
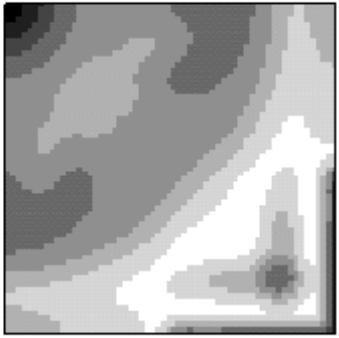
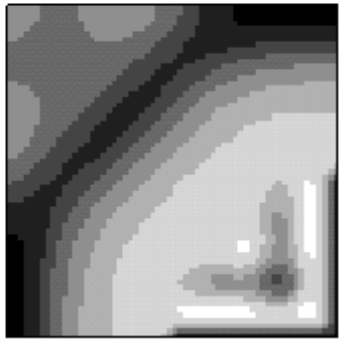
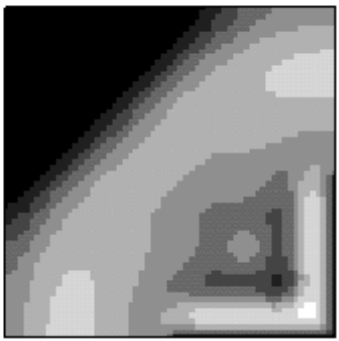
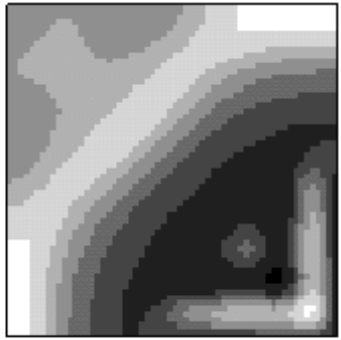
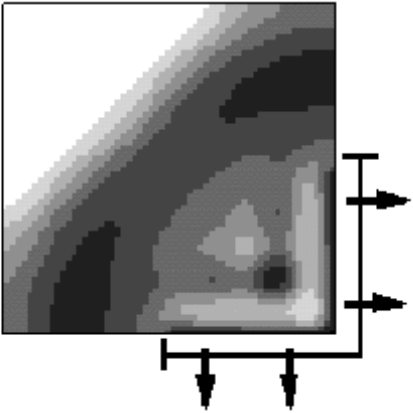
\end{document}